# Response to critiques on Observation of the Wigner-Huntington transition to metallic hydrogen.

I.F. Silvera and R. Dias
Lyman Laboratory of Physics, Harvard University, Cambridge, MA 02138

**Introduction**

We reported the first observation of metallic hydrogen (MH) in the low temperature limit at a pressure of ~495 GPa in an article published in Science (*1*). This transition was first predicted by Wigner and Huntington (**WE**) over 80 years ago (*2*) at a pressure of ~25 GPa. In recent decades it became clear that the required pressure for metallization was far greater, in the 400-500 GPa range. Until now the observation of the WE transition in diamond anvil cells (**DACs**) has been prevented by one problem: the diamonds break before a sufficiently high pressure has been achieved. This has driven the high-pressure community to improve DACs and experimental methods to understand and overcome the conditions that limited the performance of diamonds and the pressure. In our experiment, with increasing pressure, we observed a clear transition from a transparent sample of solid molecular hydrogen to an opaque black sample to a shiny reflective sample of MH, as determined by reflectance measurements. There is no doubt that MH was produced at the highest pressures. Yet there have been criticisms concerning the pressure that was achieved, the possibility that the 50 nm alumina layer, deposited on diamonds to inhibit diffusion of hydrogen, might be transformed to a metal and be responsible for the reflectance, and analysis of the reflectance. Here we respond to the criticisms posted on the condensed matter arXiv by Loubeyre, Occelli, and Dumas (**LOD**)- arXiv:1702.07192, Eremets and Drozdov (**ED**)- arXiv:1702.05125, and Goncharov and Struzhkin (**GS**)- arXiv:1702.04246.

**I. Alumina Layer**

We first deal with comments concerning the alumina layer. LOD (and ED) suggest that the alumina itself might be transformed to a metal, referencing a 2011 article by Nellis (*3*). LOD failed to keep up with the literature, as in 2015 Liu, Tse, and Nellis (*4*) reanalyzed alumina and find that it melts to a metallic liquid at 500 GPa at a temperature of 10,000 K, becoming a good metal at 900 GPa. These conditions do not exist in our DAC. They also suggest that alumina might be reduced by hydrogen to alane, $AlH_3$. Goncharenko et al (*5*) studied alane and found that it becomes a metal at 100 GPa. Subsequently Geshi and Fukazawa (*6*) have shown that at ~300 GPa alane opens up a gap and would be transparent. In our studies we see no change of the alumina at ~100 GPa, so it is not reduced. Furthermore, at the highest pressures we see clear images of the gasket and sample through the diamonds and coatings as shown in our paper (Figs. 2 and S6). We reproduce some of these in Fig. 1, below. Such images would not be possible if alumina had converted to a shiny metal.

ED also question reflective effects if the diamond culet is not flat, i.e. cupped. We measure reflectance with light normal to the culet. A straightforward geometric calculation of the maximum cupping for a culet of 30 micron diameter and a gasket that is about 1 micron thick (as in our case at high pressure) yields a radius of curvature of 225 microns with a half angle of 3.8 degrees subtending the putatively cupped gasket. Thus, there should be little impact on the measured reflectance at normal incidence. ED also state "The only direct way to establish a metallic state is to measure the electrical conductivity of the sample down to the lowest temperatures". Measurement of DC conductivity is one method. However, there are numerous

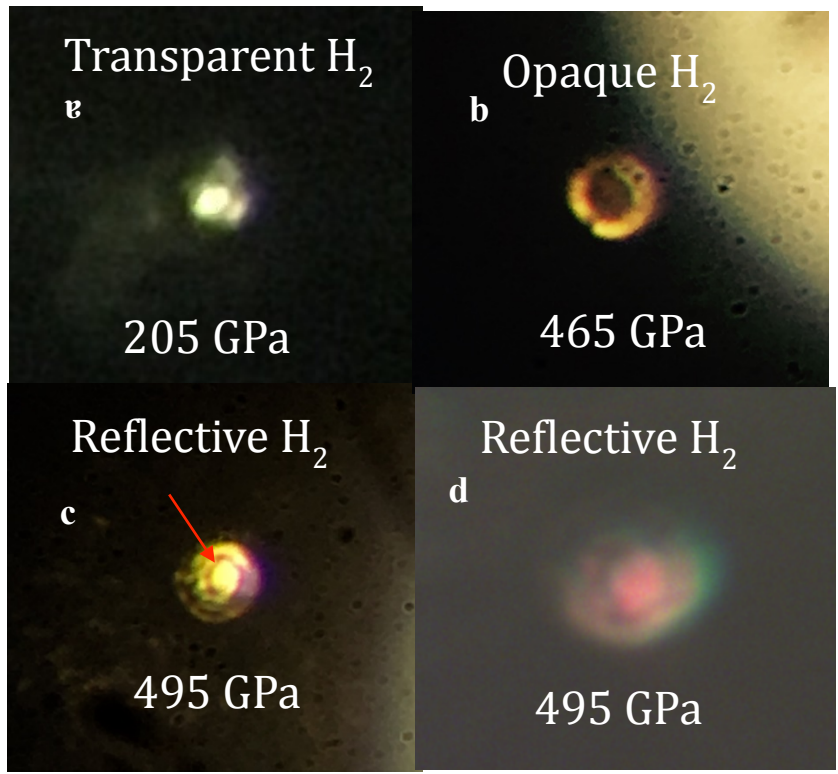

Fig. 1. Clear images of MH showing the sample and the gasket. Resolution has some limitations as the object distance of the images is ~125 mm and the cryostat has three $CaF_2$ windows between the microscope and the DAC (there was no optical correction for these windows). a,b,c) Hydrogen imaged with a stereo microscope using an iphone camera at the ocular. The boundary between the MH and rhenium gasket is very clear in c. We note that the magnification used with the iphone was not recorded and can be slightly different for the three iphone images. d) An image taken with a Wild 420 Macroscope. The image is magnified and projected onto a CMOS color camera. The color spectrum for c and d are different. The gamma function of the CMOS camera was set to 1 (gamma of 1 gives input and output intensities the same over the spectral range; gamma was the same for all iphone photos but not 1; we could not control the gamma function of the iphone, thus the colors are different in c and d)

static and dynamic studies in which metallic reflectance, which is due to the interaction of high frequency electromagnetic waves with free electrons, has been used to determine transitions to the metallic phase (*7-10*).

**Semiconducting State**
   A semiconductor has a fundamental reflectance in the low temperature limit. As the temperature is raised, electrons from the valence band populate the conduction band and these free electrons give rise to increased reflectance. Thus, when the temperature is decreased and the conduction band is depopulated, the reflectance should decrease for a semiconductor. Our reflectance measurements between 83 and 5.5 K show increased reflectance, as expected for a metal.

**II. Pressure Determination**
   The three comments are all concerned with the pressure that was given for the metallization of hydrogen. First, we make some general statements that we are sure can find agreement amongst the commenters. There is no metrological standard for determining pressure in DACs! The higher the pressure, the more challenging is the determination. The static high-pressure community has adopted the ruby fluorescence scale for pressures to ~150-200 GPa. Recent extensions to pressures of 200-400 GPa have relied on the shift of the Raman active phonon in the stressed diamond, calibrated against metallic markers (markers are pressurized metals that are X-rayed to determine the pressure and volume and compared to P-V equations of states of the metals, determined from shock wave experiments and theoretically corrected for the high temperatures of the shocked metal). It has been shown that the shift of the Raman phonon depends on geometric factors (*11, 12*) giving rise to uncertainties. Calibration data have been fit to linear as well as quadratic scales and extrapolated to higher pressures.

Considering these factors the determined pressure can have large systematic uncertainties and this was stated in our paper, where we gave the determined value of the pressure, along with the assumptions. It is important that we provided a diamond Raman spectrum corresponding to the metallization pressure. If in another experiment that spectrum is reproduced, hydrogen should be in the metallic phase.

Some commenters state that we did not continuously measure the pressure throughout the experiment with laser Raman scattering. It is well-known that shining a laser into high-pressure stressed diamonds can lead to failure of the diamonds. In our experiment we did not measure the pressure with a laser until the reflectance measurements were made, and then with a red laser and very low power of a few mW. Thus, we have a Catch-22 situation. If use a laser we may not get to the highest pressure and no MH; not using a laser one may achieve the pressure for MH, but there may be uncertainty and criticism of the highest pressure. We decided to go for the highest pressure; we made MH and then measured the pressure with the diamond Raman scale.

We comment here that we no longer have our sample of MH. We maintained it at high pressure and low temperature for four months and could not observe any visual change in reflectance or dimensions while our paper was under review. Subsequently, before carrying out new experiments we shined less than 1 mW of laser power (642.6 nm) into the diamonds to again measure the pressure and observed immediate catastrophic failure of the diamonds, as has been experienced by others in high pressure research.

**IIa. Black Hydrogen**

LOD focus attention on hydrogen when it is black, which was first observed by Loubeyre et al (*13*) at a pressure reported to be 320 GPa. They then use this pressure to redetermine our pressures and propose that pressures are much lower than we report. Some years ago the reported pressure of black hydrogen created conflict with an experiment by Narayana et al (*14*) who earlier had observed that hydrogen was transparent at a higher pressure of 342 GPa, using an X-ray marker for pressure determination. Loubeyre et al had determined their pressures by extrapolation of the Ruby pressure scale from 80 GPa (they used the calibration of Mao et al (*15*) of ruby in an argon pressurization medium that was subsequently determined to be non-hydrostatic (*16*)). This motivated a new calibration of the ruby scale to 150 GPa in a quasi-hydrostatic pressurization medium (*17*). Extrapolation of this improved scale showed that the pressure of black hydrogen given by Loubeyre et al might be 60 GPa higher than reported, or around 380 GPa. LOD now state that their pressure was even lower, 300 GPa, using the marker calibrated diamond Raman scale, although their high pressure data is not shown. They also refer to an observation of black hydrogen at ~270 GPa (*18*). By contrast Dias et al (*19*) observed black hydrogen at ~400 GPa in an earlier measurement (*19*), and again in the same pressure region in the MH paper. This is supported by other measurements. Eremets et al (*20*) report hydrogen becoming darker at 380 GPa. Also, Dalladay-Simpson do not observe black hydrogen at pressures to 380 GPa and room temperature (*21*), while Zha et al (*22*) show pictures ( Fig. 2, below) in

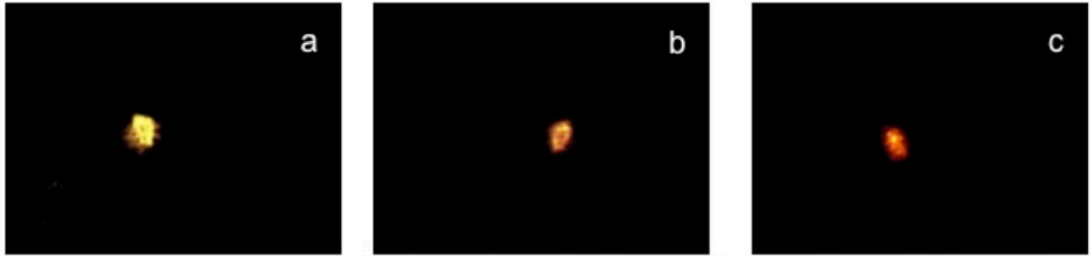

**Figure 2.** Hydrogen in transmitted light at pressures of 310, 340, and 360 GPa at 80 K in figs. a, b, c, respectively.

transmitted light indicating that hydrogen is becoming a bit darker at the highest
 Thus, we conclude that the pressure where hydrogen turns black is quite controversial. The pressure of the appearance of black hydrogen should not be used as a "fixed point" to rescale our pressures, as LOD have done.

 LOD criticize our pressure determination and present their unpublished calibration of the frequency of the IR vibron vs. pressure, compared to our calibration data, reproduced below in Fig. 3, implying that our calibration is erroneous. In Fig. 4 we replot LOD's calibration points on an extended scale, along with several other calibrations (*1, 19, 22, 23*), including our own, for comparison. There is a striking disagreement between the calibration data of LOD and the other calibrations, which all group together. An extrapolation of LOD's calibration to 150 GPa over-estimates the pressure and has a large discrepancy with the vibron frequency that is well established in the literature. The deviation from other calibrations is even greater at higher pressures.

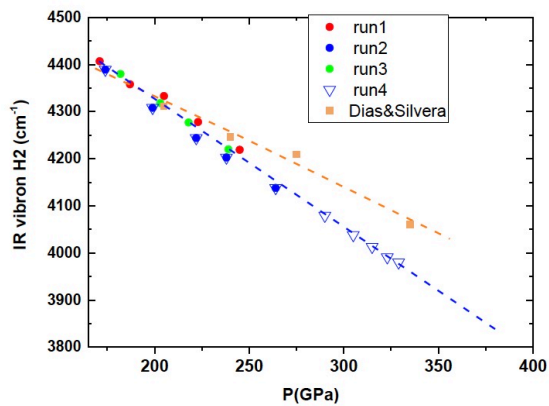

**Figure 3.** The unpublished calibration of frequency vs pressure presented in the comment by LOD.

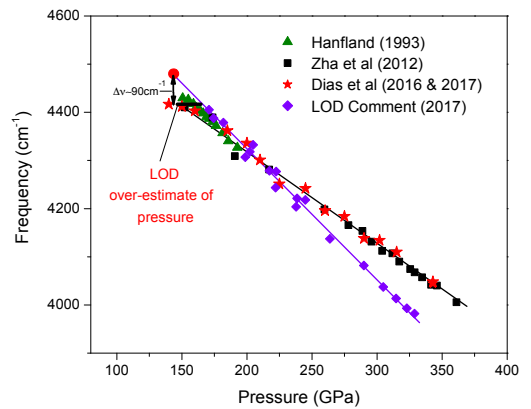

**Figure 4.** An extension of the pressure range showing the calibrations of several groups for the IR vibron frequency, including our own, compared to that of LOD.

We used calibration data from Zha et al (*22*) for determination of pressure using IR modes of hydrogen. LOD are critical of this calibration stating that "*the pressure was not directly measured but estimated using the extrapolation of lower pressure measurements*". However, Zha et al state in their SI that they used their Raman phonon data for their calibrations. They did a careful calibration of the pressure, shown in Fig.5, below. Thus, LOD apparently mistakenly misrepresent the paper of Zha et al. We conclude that unpublished calibration presented by LOD has serious deviations from several other calibrations and may possibly be in error.

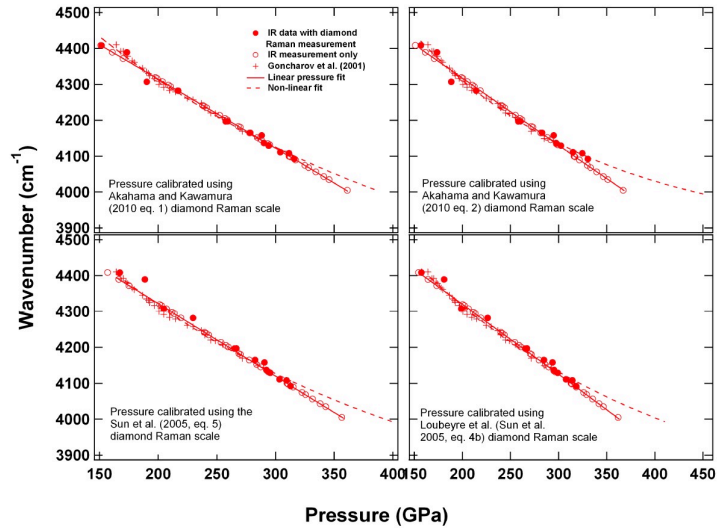

**Figure 5.** Figure 1 from the SI of Zha et al (*22*) showing their calibration data and comparing to other calibrations.

### IIc. Rotation/Load Curves

We found that our load scale was a useful secondary means for determining the pressure, since we had decided not to use lasers to measure the pressure. LOD as well as ED criticize the use of a rotation/load scale that we used for pressure. LOD attempt to recalibrate our pressures using their "fixed-point" scale of black hydrogen, which we have shown above is unreliable. Both LOD and ED show their own load scales that are non-linear. In our DACS the structure is such that diamonds are loaded against the gasket/sample without deformation of the body of the cell (*24*); it is the arms of our cell that elastically deform with load and this deformation is measured with strain gauges. The type of DACs used by LOD and ED is not clear from the literature, except that LOD use a diaphragm or membrane cell with a very different structure than ours. The data for our rotation scale, which has a rather linear behavior, are experimentally determined pressures from the IR vibron peaks that had been calibrated by Zha et al. and from the Raman phonon of stressed diamond. We pointed out in our paper that eventually our linear scale will plateau, as we have observed in earlier experiments. In Fig. 6 we show three of our rotation/pressure curves; the same DAC was used in all three experiments. It is our experience that when the rotation curves plateau, the diamonds are soon to fail, as increased load does not increase the pressure. It is interesting that for the MH run, the curve has not yet plateaued, indicating that the pressure could have been increased.

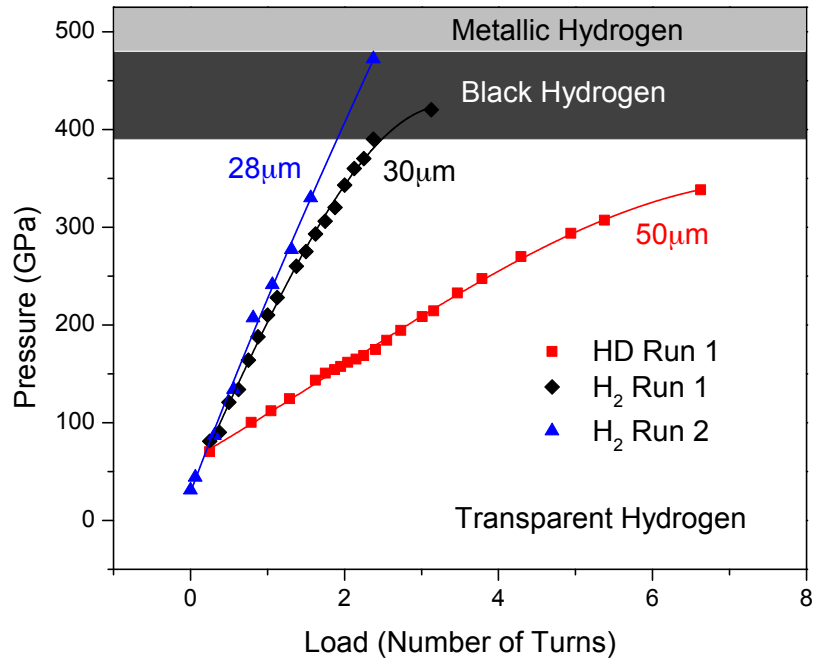

**Figure 6.** Load or rotation vs pressure for our DAC for a few different high pressure experimental runs: HD Run1 (*25*); H$_2$ Run 1(*19*); H$_2$ Run 2 (*1*). The data points are measured pressures. Each curve indicate the diameter of the diamond culets in microns.

GS comment *"Previously used extrapolations of the pressure-load curves cannot serve as a reliable tool any more: moreover, the linear curves of this kind are totally unrealistic"*. Our load curves are measured for each experimental run!

### IId. The Diamond Raman Phonon

ED and LOD comment on our diamond Raman spectrum. We have substantial experience observing such spectra; these spectra can differ from experiment to experiment and ours is not so unusual. Our spectrum exhibits a double peak (in Fig. 7 we reproduce Fig. S2 from our article). The triply degenerate phonon of unstressed diamond splits into a singlet and doublet as the degeneracy is lifted. Such a splitting has also been observed by others (Figs. 7, 8). Since the Raman spectrum is an important issue we show other observations in Fig. 8. Some spectra only show a falling slope at the high frequency edge such as in Fig. 8c, 206 GPa. The rise of intensity at the leading edge, such as in other Raman spectra that we show has been attributed to the quasi-hydrostatic nature of hydrogen. The large zero-point motion of atomic hydrogen, as well as the possible liquid state leads to quasi-hydrostatic behavior of the hydrogen, and may be responsible for the shape of our spectrum.

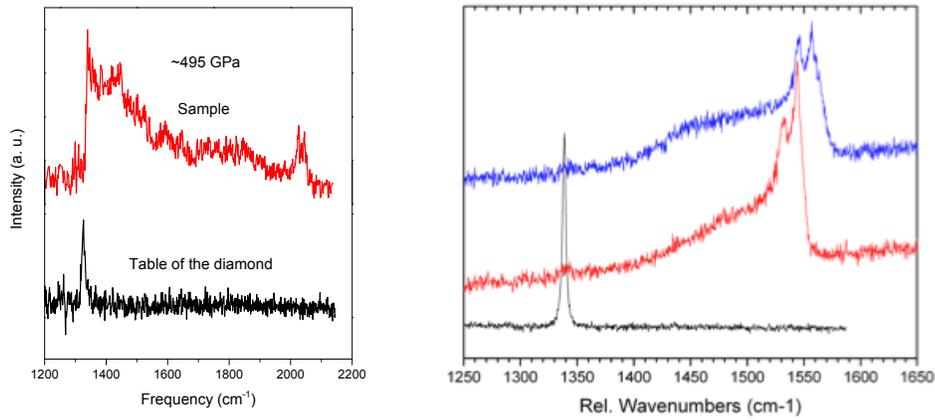

**Figure 7.** Left: Raman spectrum of diamonds in our MH experiment showing a splitting of the Raman peak. Right: A Raman spectrum from Salamat (unpublished) showing splitting of the Raman peak. In this case the incident laser beam was at a large angle to reduce normal incidence spectra arising from the less stressed regions of the diamond.

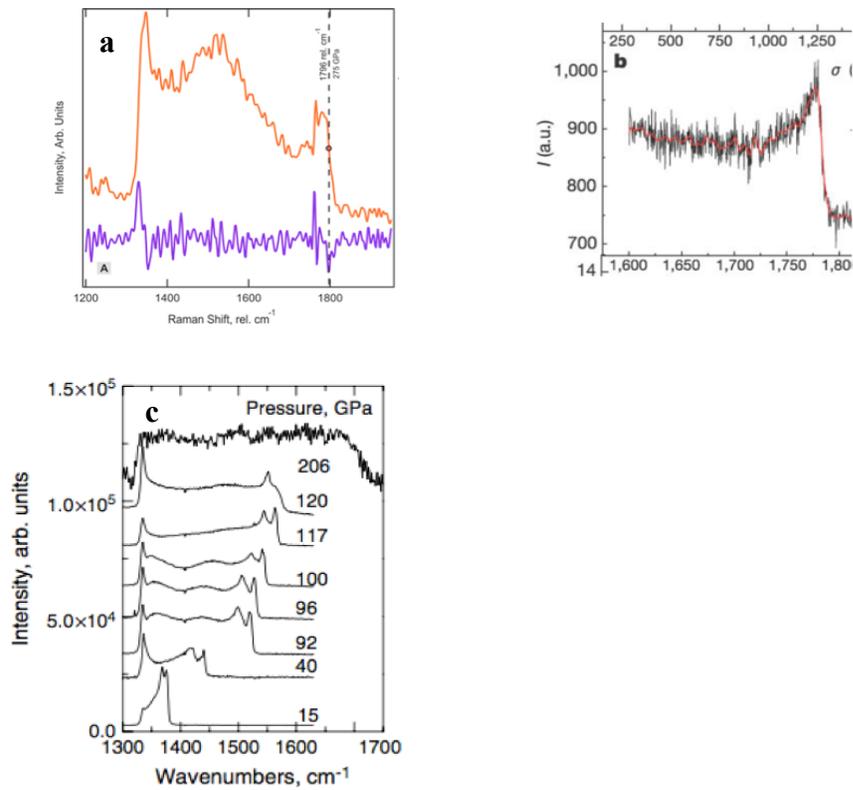

**Figure 8.** Spectra of the Raman active phonon in stressed diamond. a) From Dalladay et al (*21*), b) From Loubeyre et al (*13*), and c) From Eremets (*26*).

**IIe. Ruoff's Empirical Scale**

LOD also comment that our pressure violates Ruoff's empirical relation for the maximum achievable pressure [P(GPa)= 1856 D(μm)$^{-1/2}$, D is the diameter]. This expression states the smaller the culet diameter the higher the pressure. Ruoff's empirical law works in some cases and not in others, and thus should be regarded as a rule-of-thumb, only. The general experience in high-pressure physics is that sometimes diamonds fail at very high pressures and sometimes at unexpectedly low pressures. Ruoff's law has many exceptions and we give some examples in Table I, below.

Some years ago one of us (IFS) and A. Ruoff tried to understand why higher pressures are generally achieved with smaller culet diamonds. We concluded that the failure of diamonds may be due to defects in the culet flat and the larger the flat the higher the probability of a defect being included. We (IFS) started examining culets for microscopic defects with atomic force microscopy-AFM (see Fig. S7 of our article). The culets of these diamonds looked beautiful under bright field microscopy, but defects could be observed using AFM. Based on these observations we started a program to etch off several microns from the surface of the diamond anvil culets, using reactive ion etching (RIE). Our success in achieving higher pressures with large culets may in part be due to this procedure.

| Culet Diameter (μm) | Ruoff's proposed law P=1856/D(μm)$^{-1/2}$ (GPa) | Max. Pressure achieved Experimentally (GPa) and Reference | Validity of law Percent difference |
|---|---|---|---|
| 200 | 131 | **201** Drozdov et al Nature 525(2015) | Not Supported 153% |
| 180 | 138 | **202** Dias et al PNAS 110 (2013) | Not Supported 146% |
| 100 | 185 | **220** Yoo et al Angew. Chem. Int. Ed. (2011) | Not Supported 119% |
| 50 | 262 | **340** Dias et al PRL, 116 (2013) | Not Supported 130% |
| 30 | 338 | **365** https://arxiv.org/abs/1702.05125 (2017) | Not Supported 108% |
| 30 | 338 | **420** Dias et al https://arxiv.org/abs/1603.02162 (2016) | Not Supported 124% |
| 28 | 350 | **495** Dias et al Science, 335 (2017) | Not Supported 141% |

**Table I.** Examples of violations of Ruoff's empirical formula.

**III. Reflectance.**

An important part of our observations is the measurement of the energy dependence of the reflectance. We use normal incidence reflectance and measure the reflectance from the table of the diamond for normalization, as had been done in earlier studies (*27*). We have already dispensed with the critique of the alumina, above. As shown in. 1 of our paper, the sample transformed from black, light absorbing hydrogen, to a shiny reflective sample manifest of a metallic state. The uncorrected reflectance in the infrared (1550 nm) and red (642.6 nm) was from ~0.90 to 0.85, respectively. However, the raw reflectance in the green and blue region falls off substantially. This is associated with absorption of light as it is transmitted through the stressed region of the diamond. In our analysis we applied results for absorption from Vohra (*28*) and the critique of GS raises an important point. The high-energy absorption of diamond is complex and in need of further study. The data from Vohra may not be applicable for the diamonds that we used: CVD (chemical vapor deposition) grown type IIac diamonds (supplied by Almax-Easylab). Although the Vohra data implied highly attenuated light in the blue, we could measure substantial reflectance in this part of the spectrum. It has been believed that the attenuation in the UV-blue region of the spectrum was due to a closing of the energy gap in diamond (*29*), supported by theoretical calculations (*30*). However, a recent experimental study at SLAC (*31*), using shock wave and X-ray techniques reports that the band gap of diamond opens with increasing pressure. If this is the case then the UV-blue absorption observed in stressed diamond in DACs may be due to impurities or induced defects, and this could depend strongly on the quality of the diamonds before being stressed. We have refit our reflectance (Fig. 9) using only measurements in the red and infrared where the diamonds are likely transparent. The resulting values of the plasma frequency (33.2±3.5 eV) and scattering time (5.9±0.8x10$^{-16}$ sec) do not change any of the conclusions in our paper.

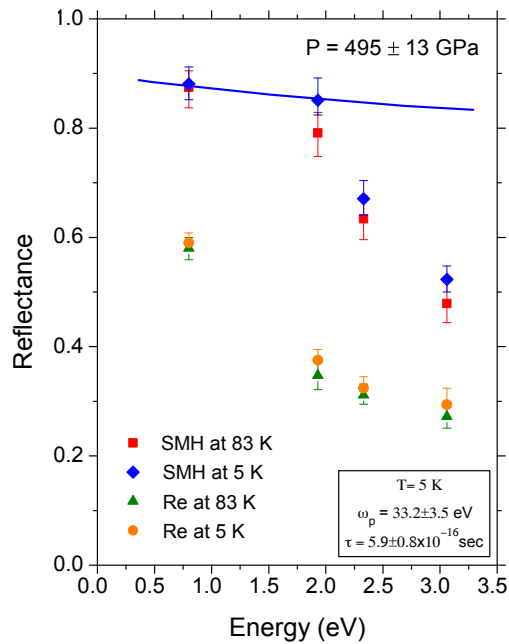

**Figure 9.** The reflectance of MH fit only to the red and infrared reflectance measurements without correction for the diamonds. Reflectance from the diamond table was used for the reference intensity.

**IV. Discussion**

In order to achieve higher pressures we implemented a number of procedures:
- Use of synthetic diamonds
- Removal of surface defects by RIE
- High temperature annealing of diamonds
- Precision alignment of diamond anvils for very high loads
- Coating diamonds with alumina that acts as a diffusion barrier
- Maintain the hydrogen/diamond cell at low temperatures at all times to inhibit diffusion
- Never shine laser light with powers above several milliwatts into the stressed diamonds. This can lead to weakening of the diamonds or failure.

We are not sure which of these procedures enabled us to achieve the high pressures. We point out that perfect diamonds should have the strength to perform to higher pressures. The fact that they fail at lower pressures must point to imperfections in the diamonds. Some of these are visible and can be removed, or diamonds can be selected for certain properties. However, there remain imperfections that are difficult to identify, so we maintain a focus on the diamonds to achieve the highest pressures.

We hope that this paper has further clarified the issues that were brought up by the LOD, ED, and GS and that skepticism can be converted into efforts to further understand the properties of MH.

Acknowledgements: We thank Mohamed Zaghoo, Rachel Husband, Ashkan Salamat, Ori Noked, Bill Nellis, and Marius Millot for comments and discussions. The NSF, grant DMR-1308641 and the DoE Stockpile Stewardship Academic Alliance Program, grant DE-NA0003346, supported this research.